\documentclass[aps,twocolumn,preprintnumbers,amsmath,amssymb]{revtex4-2}

\usepackage{graphicx}
\usepackage{dcolumn}
\usepackage{bm}
\usepackage{multirow}
\usepackage{color}
\usepackage{ulem}
\usepackage{amsmath}
\usepackage{gensymb}

\begin{document}
\title{Efficient valley polarization of charged excitons and resident carriers in MoS$_2$ monolayers by optical pumping}

\email{fabian.cadiz@polytechnique.edu}
\author{S. Park $^{1}$}
\author{S. Arscott $^{2}$}
\author{T. Taniguchi$^3$}
\author{K. Watanabe$^4$}
\author{F. Sirotti $^{1}$}
\author{F. Cadiz$^{1}$}

\affiliation{$^1$ Laboratoire de Physique de la Mati\` ere Condens\' ee, CNRS, Ecole Polytechnique, Institut Polytechnique de Paris, 91120 Palaiseau, France}

\affiliation{$^2$ University of Lille, CNRS, Centrale Lille, Univ. Polytechnique Hauts-de-France, UMR 8520-IEMN, F-59000 Lille, France}

\affiliation{$^3$International Center for Materials Nanoarchitectonics, National Institute for Materials Science, 1-1 Namiki, Tsukuba 305-0044, Japan}

\affiliation{$^4$ Research Center for Functional Materials, National Institute for Materials Science, 1-1 Namiki, Tsukuba 305-0044, Japan}


\begin{abstract}
We investigate with polarized microphotoluminescence  the optical pumping of the valley degree of freedom in charge-tunable MoS$_2$ monolayers encapsulated with hexagonal boron nitride at cryogenic temperatures. We report a large steady state valley polarization of the different excitonic complexes following circularly-polarized laser excitation $25$ meV above the neutral exciton transition. For the first time in this material we reveal efficient valley pumping of positively-charged trions, which were so far elusive in non-encapsulated monolayers due to defect and laser-induced large electron doping. We find that  negatively-charged trions present a polarization of $70\%$ which is unusually large for non-resonant excitation. We attribute this large valley polarization to the particular band structure of MoS$_2$, where an optically dark exciton ground state coexists with  a bright conduction band ordering in the single-particle picture, leading to a supression of the valley relaxation for negatively-charged trions. In addition, we demonstrate that circular excitation induces a dynamical polarization of resident electrons and holes, as recently shown in tungsten-based monolayers. This manifest itself as a variation in the intensity of different excitonic complexes under circular and linear excitation.
\end{abstract}


\maketitle
\textit{Introduction.---}
Two-dimensional crystals of transition metal dichalcogenides (TMD) such as MX$_2$ (M=Mo, W; X=S, Se, Te) have emerged as promising, atomically-thin semiconductors for applications in valley/spintronics \cite{behnia:2012,Xiao:2012a,Butler:2013a}.  Indeed, the interplay between inversion symmetry breaking in monolayers (ML) and the strong spin-orbit interaction induced by the heavy transition metal atoms yields a unique spin/valley coupling which is expected to provide additional functionalities in future devices \cite{Xiao:2012a,Li:2020a,Li:2020b,Huang:2020a}. The absence of inversion symmetry in the single layer limit is also at the origin of chiral optical selection rules which enables an all-optical manipulation and readout of both the  valley and spin degree of freedom. Due to enhanced Coulomb interaction in 2D, weak dielectric screening and large effective masses, the optical excitation couples mostly to exciton and trion resonances \cite{Ramasubramaniam:2012a,Ross:2013a,Mak:2013a}. Remarkably, light absorption due to these strongly bound excitons preserves the coupling between light chirality and the valley.
 The very first demonstrations of a valley polarization controlled by light helicity in TMDs were obtained using monolayer MoS$_2$  \cite{Cao:2012a,Mak:2012a,Zeng:2012a,Sallen:2012a,Kioseoglou:2012a}, the first member of the TMD family to be isolated into a single layer and shown to become a direct bandgap semiconductor- in contrast with the indirect bandgap character of bulk MoS$_2$ \cite{Splendiani:2010a, Mak:2010a}. The important role played by MoS$_2$ in early studies is explained by the high  abundance of the naturally occurring mineral molybdenite \cite{Dickinson:1923a}. However in the following years after these key experiments, detailed optical studies on the optical manipulation of the valley degree of freedom  focused mainly on monolayer WSe$_2$ and MoSe$_2$ due to their apparent superior optical quality with respect to MoS$_2$.  Importantly, in these two materials the application of gate voltages allowed one to tune the resident carrier density and therefore to study the valley properties of neutral, positively and negatively charged excitons.
  In contrast, non-encapsulated MoS$_2$ suffers from large inhomogeneous broadening and strong photo-induced irreversible changes in the photoluminescence (PL) emission after laser exposure \cite{Cadiz:2016b,Wierzbowski:2017a}. As a result, the PL spectrum of MoS$_2$ at low temperatures was systematically dominated by negatively charged excitons and electrostatic doping in the hole regime was never achieved. Only very recently electrostatic p-doping was achieved in MoS$_2$ monolayers encapsulated with hexagonal boron nitride (h-BN) \cite{Klein:2021a}.  This encapsulation has been shown to dramatically improve the PL linewidth in TMDs and also prevent unintentional electron doping in MoS$_2$ \cite{Cadiz:2017a}.  The high optical quality provided by h-BN encapsulation had open the door for the observation of dark excitons \cite{Robert:2017a,Liu:2019a,Zinkiewicz:2020a}, phonon-replicas \cite{Liu:2019b,Robert:2021b}, donor-bound excitons \cite{Rivera:2021a}, neutral and charged bi-excitons \cite{Chen:2018a,Barbone:2018a,Ye:2018a} and, more recently, the well-resolved fine structure of negatively-charged excitons in tungsten-based TMD monolayers \cite{Courtade:2017a,Vaclavkova:2018a,Lyons:2019a}. This trion fine structure has been recently used to demonstrate efficient valley polarization of resident electrons in WSe$_2$ and WS$_2$ monolayers\cite{Robert:2021a}.  In this work, we have fabricated a charge-tunable device with encapsulated MoS$_2$; and we present the first investigation of the steady-state valley polarization of both excitons and resident carriers under circularly-polarized excitation in all doping regimes. \\

\begin{figure*}
\includegraphics[width=0.85\textwidth]{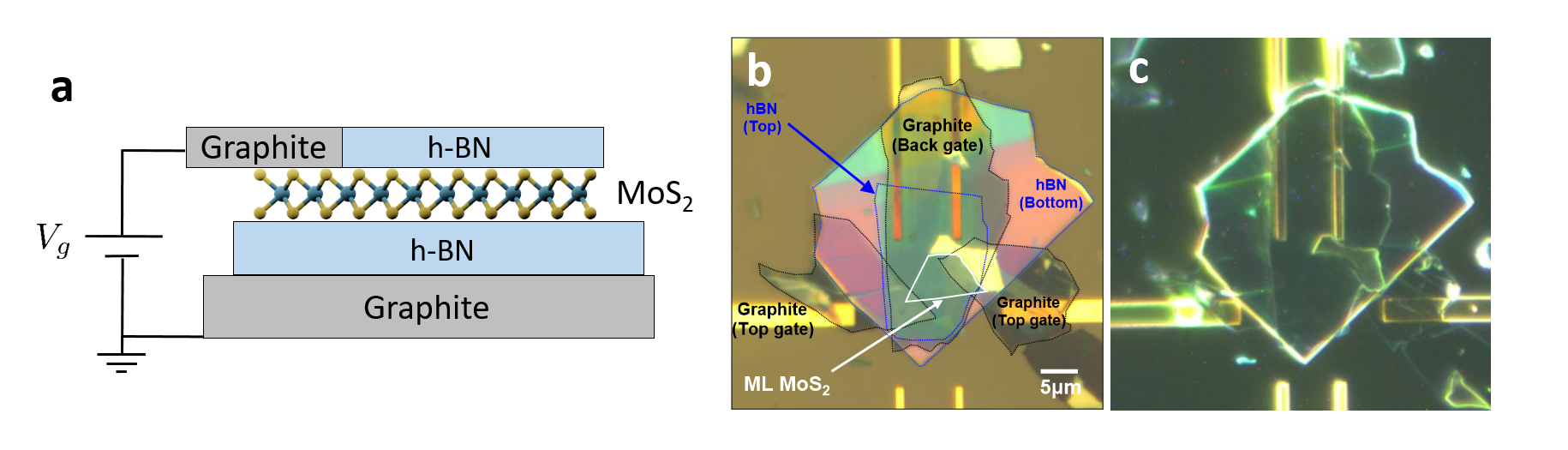}
\caption{\label{fig:fig1} (a) Schematic side-view of the sample. The MoS$_2$ monolayer is encapsulated between two thin h-BN flakes to provide high optical quality and to prevent photodoping effects. Thin graphite flakes were used to provide an electrical control of the resident carrier concentration in the monolayer. The whole heterostructure is deposited onto a silicon substrate with a 90 nm-thick silicon dioxide  layer on top of which gold lines were deposited lithographically prior to the mechanical exfoliation. (b) Optical microscope image of the sample under white light illumination. (c)   Dark-field image of the sample, revealing clean interfaces with no visible bubbles after the formation of the heterostructure.}
\end{figure*}

\indent \textit{Samples and Experimental Set-up.---}
A charge tunable device such as the one shown in Fig.\ref{fig:fig1} (a) was fabricated by mechanical exfoliation of bulk MoS$_2$ crystals from 2D semiconductors. Exfoliation was done inside a glovebox where oxygen and water levels are kept below 1 ppm.  100nm Au/5nm Ti lines were pre-patterned onto an SiO$_2$ (90 nm)/Si substrate using lithographic techniques and physical vapor deposition onto which the heterostructure was then deposited by using a transparent viscoelastic stamp \cite{Gomez:2014a}. Two graphite flakes were used to contact the MoS$_2$ monolayer to one of the gold electrodes which permits for an electrostatic doping by the application of a gate voltage $V_g$. \\

\noindent
A hyperspectral confocal  micro-PL ($\mu$PL) set-up is used to excite and detect the polarized exciton emission at cryogenic temperatures \cite{Favorskiy:2010a, Cadiz:2018a}. The samples are kept inside a closed cycle He cryostat and excited with a continuous wave solid-state laser at 633 nm ($\sim 1.96$ eV). The polarization of both the laser and the detected PL is controlled with liquid crystal retarders and linear polarizers. 
The  sample temperature is maintained at $20$ K. The laser beam is focused onto a diffraction-limited spot (of Gaussian radius $\approx 0.4\;\mu$m) in the sample plane thanks to a vacuum (and cryogenic) compatible, apochromatic objective mounted inside the cryostat (numerical aperture NA= 0.82).
  The resulting PL spot is imaged onto the entrance slit of a $320$ mm focal length spectrometer 
 equipped with a 600 grooves/mm diffraction grating. For micro-reflectivity measurements, an image of a pinhole (diameter 50 $\mu$m) illuminated by a quartz-tungsten halogen lamp is formed by the objective into the sample plane, thus probing  the reflectivity of the monolayer over a typical diameter of $\sim 1\;\mu$m. \\

\indent \textit{Results and Discussion.---}
Fig.\ref{fig:fig1} (a) shows a schematic drawing of the charge tunable Van der Waals heterostructure deposited onto a SiO$_2$/Si substrate containing lithographically-defined gold connections. A microscope image of the sample under white light illumination is shown in Fig.\ref{fig:fig1}(b), whereas the dark field image of the same region is shown in Fig.\ref{fig:fig1}(c). The absence of visible bubbles and wrinkles in the latter image is a signature of the clean interfaces achieved during the transfer process\cite{Purdie:2018a,Jain:2018a}. This is confirmed by the well-resolved features observed in the PL spectrum at low temperatures.\\

\begin{figure*}[htbp]
\includegraphics[width=1.05\textwidth]{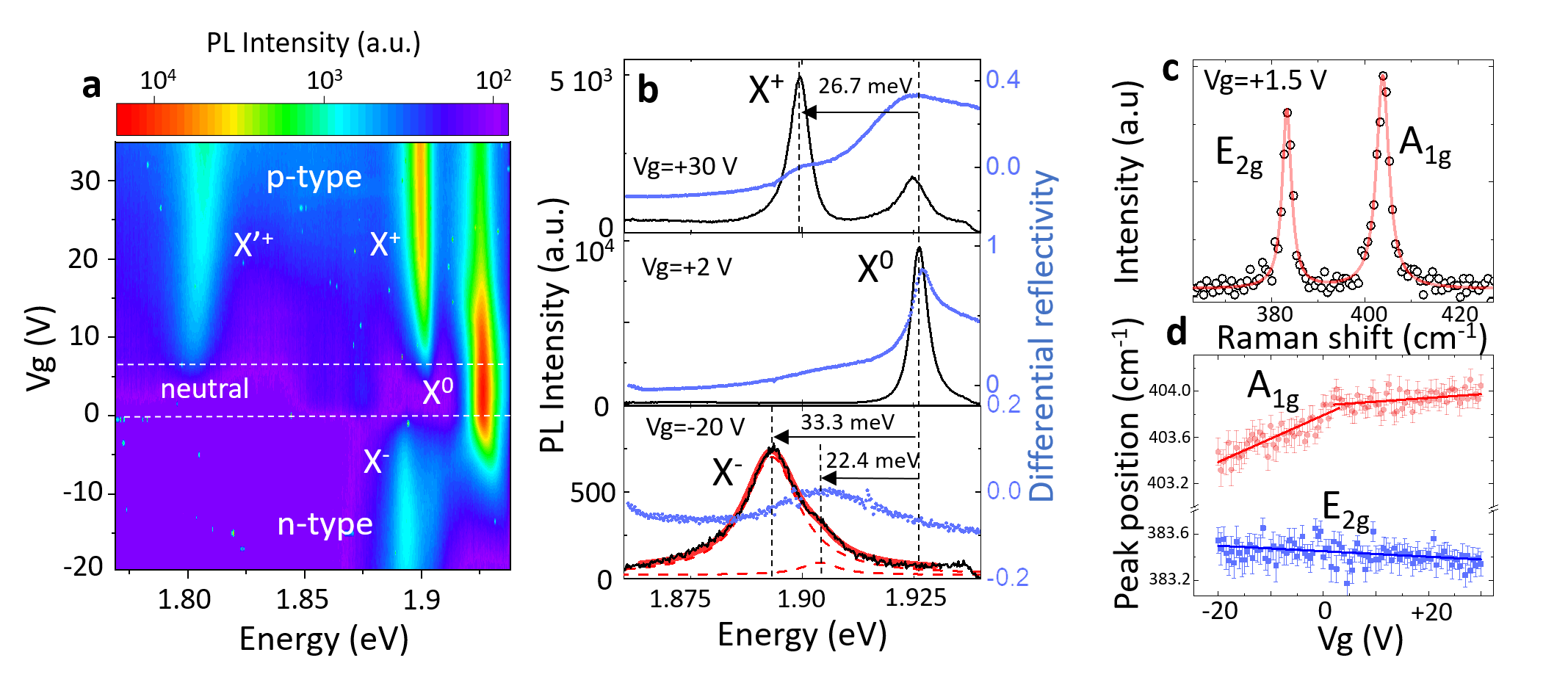}
\caption{\label{fig:fig2} (a) PL intensity on a logarithmic scale under cw laser excitation at $10\;\mu$W as a function of photon emission energy and gate bias (b) PL spectra at selected gate voltages corresponding to the p-type (top), neutral (middle) and n-type doping regime (bottom).  Also shown is the differential reflectivity spectrum (filled dots) for each case.  (c) Raman spectrum in the neutral regime, exhibiting the well-known E$_{2g}$ and A$_{1g}$ vibrational modes. (d) Peak position for both modes of panel (c) as a function of gate bias.}
\end{figure*}

\noindent
As shown previously, encapsulation by  h-BN provides linewidths close to the homogeneous limit at low temperatures  \cite{Cadiz:2017a,Jakubczyk:2016a,Martin:2020a}, allowing for a clear spectral separation of the different excitonic complexes and the study of their valley polarization under circularly-polarized excitation. Moreover, encapsulation allows one to achieve both n and p-doping by the application of a gate voltage, so far inaccessible in non-encapsulated MoS$_2$. Recent scanning tunneling spectroscopy (STS) measurements have shown that h-BN encapsulation moves the Fermi level of monolayer MoS$_2$ closer to the middle of the electronic bandgap, significantly reducing the electron doping when compared to monolayers exfoliated onto SiO$_2$   \cite{Klein:2019a}. This is probably due to the fact that h-BN encapsulation inhibits charge transfer between the monolayer and the SiO$_2$, as well as with atoms and molecules that may act as molecular gates when interacting with native defects. Successful p- and n-type doping in our device is demonstrated in Fig.\ref{fig:fig2}(a), where the PL spectrum is plotted against the gate voltage under a 10 $\mu$W cw excitation. Both positive ($X^+$) and negatively ($X^-$) charged trions emerge at positive and negative values of $V_g$, respectively. Fig.\ref{fig:fig2}(b) shows the PL spectrum at selected voltages together with the differential reflectivity spectrum, defined as $(R_{ML}-R_0)/R_{0}$, where $R_{ML}$ is the reflected spectrum obtained when illuminating the monolayer with a white light source and $R_0$ is the reflected spectrum when illuminating all the layers except for the MoS$_2$. In both PL and reflectivity we observe efficient transfer of oscillator strength from neutral to charged excitons when changing the gate voltage.  In the n-type doping regime, the fine structure of the negatively-charged trion manifests itself as the appearance of a double peak \cite{Klein:2021b,Grzeszczyk:2021a,Jadczak:2021a}, which we resolve despite the significant broadening of the emission. The solid red line in the bottom panel of  Fig.\ref{fig:fig2}(b) is a fit with two Lorentzians, the dashed lines represent each of the two peaks separately. Their binding energies are found to be $22.4$ and $33.3$ meV, whereas the binding energy of the $X^+$ is $26.7$ meV. The n-type regime is also accompanied by a 10-fold reduction of the PL yield.  This reduction of the PL yield is consistent with recent observations \cite{Lien:2019a}, although performed at room temperature.  This is not observed in the p-type doping regime, where a moderate broadening is observed but the PL yield is not significantly reduced. \\

\noindent
A possible explanation for this asymmetry in the PL yield is the following: in as-exfoliated  MoS$_2$ monolayer, it has been observed by scanning tunneling spectroscopy and transport measurements  that sulfur vacancies create localized in-gap states that act as hole traps \cite{Vancso:2016a,Ponomarev:2018a}. In the n-type doping regime, these localized states can contribute to a fast non-radiative decay channel for the negatively-charged trions.
  In contrast, in the p-type doping regime these traps are neutralized by capturing holes. This can also explain the emission observed at $\sim 1.8$ eV and denoted as $X^{'+}$, which probably stems from positive trions in which the recombining hole is bound to one of these traps. Finally, this hole trapping can also explain why in the p-type  regime the neutral exciton emission never disapears completely, even at the largest voltages applied. Trapping of holes can indeed limit the maximum density of positive trions that we can achieve in our device. In order to have a good estimate of the density of injected electrons and holes, we have performed gate-dependent Raman spectroscopy on the same sample. Fig.\ref{fig:fig2}(c) shows a Raman spectrum in the neutral regime, where the optically active modes $E_{2g}$ and $A_{1g}$ are clearly visible. Electron (hole) doping causes a redshift (blueshift) of the $A_{1g}$ mode due to change of electron-phonon interactions, whereas $E_{2g}$ is almost unaffected \cite{Iqbal:2020a}. This is confirmed in Fig.\ref{fig:fig2}(d) which shows the position of both peaks as a function of the gate voltage.  In the n-doping regime, the A$_{1g}$ mode redshifts by $0.5\;\mbox{cm}^{-1}$ at $V_g=-20$ V, which corresponds to a maximum areal electron density of $n\sim 1.5 \times 10^{12}\;\mbox{cm}^{-2}$\cite{Chakraborty:2012a}, consistent with the absence of the low energy feature observed in \cite{Klein:2021b} which only appears above a density of $4 \times 10^{12}\;\mbox{cm}^{-2}$. Note that the A$_{1g}$ peak shifts with a reduced slope in the p-doped regime, indicating that the hole density achieved at positive $V_g=+30$ V is probably of the order of $p\sim 3 \times 10^{11}\;\mbox{cm}^{-2}$. Importantly, no significant hysteresis is observed in the behavior of the PL when varying the gate voltage. The neutrality point, defined as the value of $V_g$ for which the $X^0$
intensity is maximized always occurs at voltages between $2$ and $3$ V. No change in the PL spectrum is observed after laser exposure, excluding the presence of laser-induced photodoping effects  \cite{Cadiz:2016b}.\\

\begin{figure*}[htbp]
\includegraphics[width=1.05\textwidth]{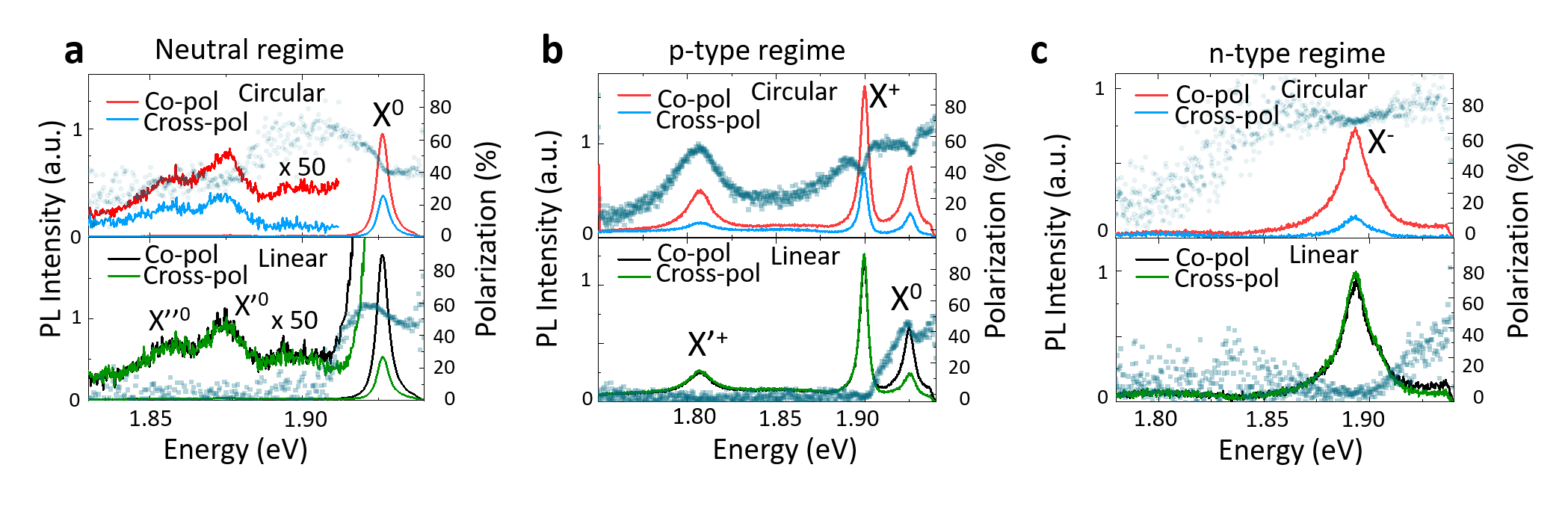}
\caption{\label{fig:fig3} (a) Polarization-resolved PL emission under circularly polarized excitation (left) and linear excitation (right) and for different gate voltages,  $+30 $V(top), $+3 V$(middle) and $-20$ V (bottom). Also shown is the energy-resolved degree of PL polarization. 
}
\end{figure*}

\noindent
In Fig.\ref{fig:fig3} we show the polarization-resolved PL intensity in the different regimes following circularly polarized excitation  or linearly polarized excitation  at $10\;\mu$W. Also shown is the degree of polarization at each emitted photon energy, defined as $\mathcal{P}= (I_{\mbox{co}}-I_{\mbox{cross}})/(I_{\mbox{co}}+I_{\mbox{cross}})$ with $I_{\mbox{co}}$ ( $I_{\mbox{cross}}$) the PL itensity co-linearly (crossed) polarized with respect to the laser. Under circular excitation, all the excitonic complexes exhibit high circular polarization, co-linearly polarized with respect to the laser indicating an efficient pumping of the valley degree of freedom in all cases. The degree of circular polarization is similar for $X^0$,  $X^+$ and $X^{'+}$, close to 50$\%$ with no strong bias dependence. 
Surprisingly, the circular polarization of the negatively charged trion complex is significantly higher, around $70\%$, consistent with some of the very first optical pumping experiments performed in non-encapsulated MoS$_2$ where a very high steady-state circular polarization of the PL was observed with a He-Ne laser excitation. In early work this high polarization was attributed to the neutral exciton emission, but it has been shown since that photodoping effects lead to heavily electron doped MoS$_2$ and total quenching of the neutral exciton on silicon substrates. So, those high values of valley polarization agree with what we observe for the negative trion, and the red He-Ne lasers used at the time were in fact practically unintentionally resonant with the $X^0$. In the neutral regime, in addition to the neutral exciton peak ($X^0$),  two peaks are clearly visible at lower energies, as shown in  Fig.\ref{fig:fig3}(a) and labelled $X^{'0}$ and $X^{''0}$. They lie at $\sim$ 51.3 meV and $\sim$ 69.5 meV below the neutral exciton, respectively, regardless of the laser excitation energy.  Since their intensities follow the same trend as the $X^0$ when varying $V_g$ (as can be seen in Fig.\ref{fig:fig1}(a)), and due to their polarization properties, we tentatively ascribe them to the recombination of $X^0$ trapped by impurities.\\

\begin{figure*}[htbp]
\includegraphics[width=0.95\textwidth]{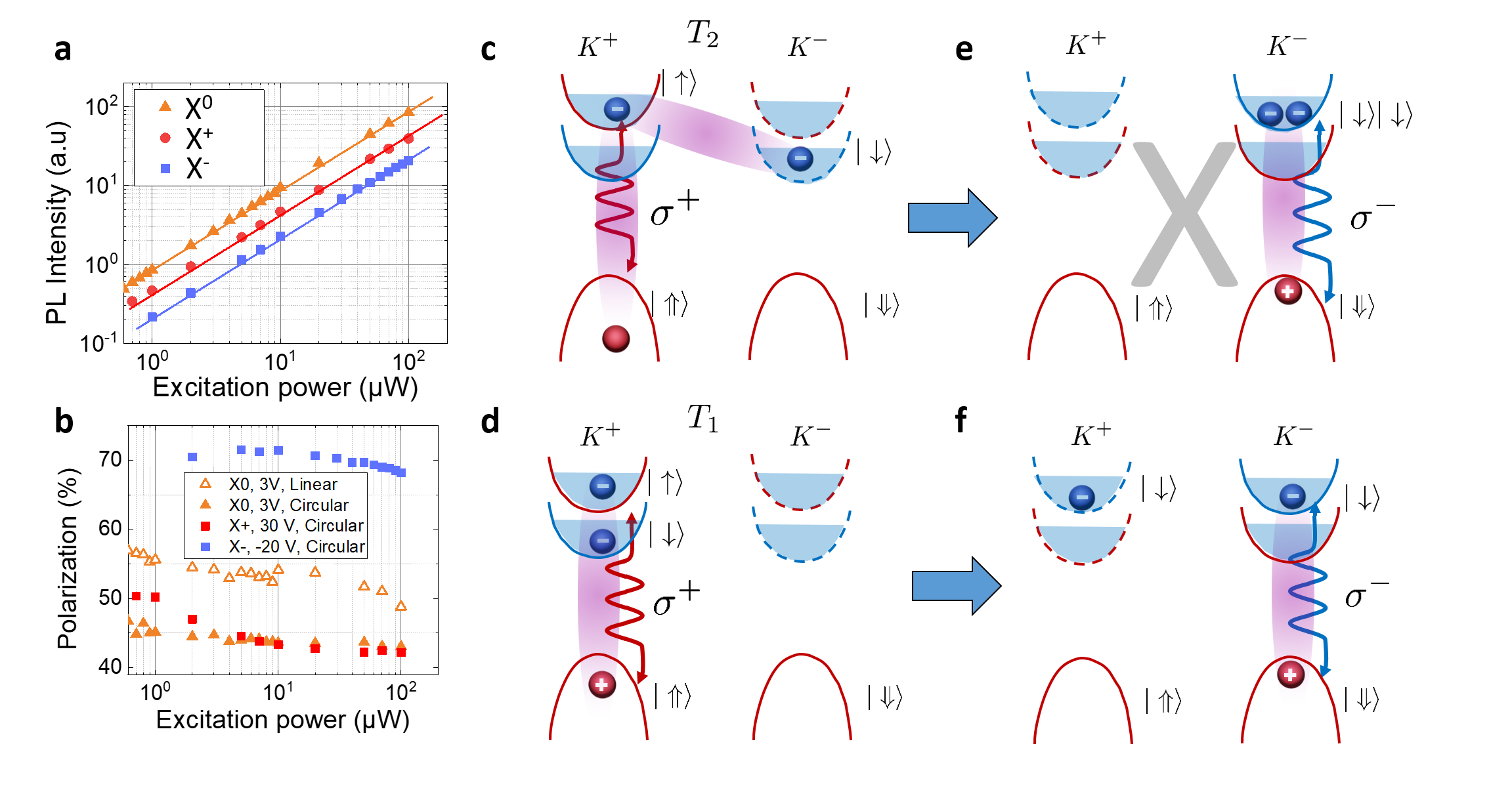}
\caption{\label{fig:fig4} (a) Integrated intensity as a function of excitation power. (b) Degree of polarization as a function of excitation power.(c) Schematics of a negatively charged trion in the $K^+$ valley corresponding to a spin singlet state in which both electrons reside at opposite valleys. (d) Negatively charged trion in the $K^+$ valley corresponding to a spin singlet state in which both electrons reside in the same valley $K^+$. 
(e) State obtained after moving the exciton from the $K^+$ to the $K^-$ valley for the trion depicted in (c). This state is forbidden by the Pauli principle.   (f) State obtained after moving the exciton from the $K^+$ to the $K^-$ valley for the trion depicted in (d).  The continuous (dashed) lines represent the ordering of the conduction bands in the excitonic (single-particle) picture. 
}
\end{figure*}

\noindent
 When exciting with a linearly polarized laser, a coherent superposition of excitons in the $K^+$ and $K^-$ valley is generated. Due to the very short exciton lifetime, in the $10^{-12}$s range \cite{Zhucr:2014a,Yan:2015b,Wang:2014b,Glazov:2015a,Robert:2016a}, this valley coherence is partially preserved before radiative recombination, and so the neutral exciton emission is co-linearly polarized with respect to the laser, with a degree of linear polarization as high as $60\;\%$. For all the other peaks, no significant linear polarization is detected. This includes the two additional peaks $X^{'0}$ and $X^{''0}$ observed in the neutral regime. If they were simply phonon replicas of the $X^0$, they should exhibit linear polarization as well.
  We have varied the excitation power by several orders of magnitude between $0.1$ and $100\;\mu$W. 
Fig.\ref{fig:fig4}(a) shows that in this power range all the excitonic species are in the linear regime, the continuous line representing a linear relationship between the integrated intensity and the excitation power. In this same power range, the degree of circular polarization is rather stable for all the complexes, but the valley coherence shows a significant drop above several tens of $\mu$W, as shown in Fig.\ref{fig:fig4}(b).\\

\noindent
We will now focus in the very high circular polarization of the negatively charged trion, which is unusual for non-resonance excitation. To explain this, we emphasize that MoS$_2$ has a unique band structure with respect to other members of the TMD family. Indeed, the small spin-orbit splitting of the conduction band is such that the spin ordering of the conduction subbands in each valley is reversed by interactions in the exciton picture. The upper conduction band having a heavier effective mass, together with the repulsive electron-hole coulomb exchange interaction leads to a "dark" configuration of the lowest energy exciton in  monolayer MoS$_2$, despite the "bright" arrangement of the spin-polarized conduction bands in the single particle picture \cite{Marinov:2017a, Liu:2013a}.  This dark character of the lowest energy exciton has been confirmed recently in magneto-optic experiments \cite{Robert:2020a}. This, together with the small spin-orbit splitting of the conduction band, makes the formation of two spin-singlet and
one spin-triplet negatively-charged trions possible-at least in principle. In a previous work  it has been argued that the triplet configuration is unbound and therefore contributes to the high energy tail of the $X^0$ \cite{Roch:2019a}. This is consistent with the doublet structure observed for the $X^-$ in Fig.\ref{fig:fig2}(b), where no third peak is clearly visible. The two peaks observed in our experiments would therefore correspond to the two spin singlets $T_{1}$ and $T_2$, whose configurations in the $K^+$ valley (coupled to $\sigma^+$ light) are represented in Fig.\ref{fig:fig4}(c) and Fig.\ref{fig:fig4}(d), respectively. Since the lowest conduction subband is more likely to be populated at moderate doping densities, the formation of the $T_2$ is more likely than that of $T_1$ under light excitation, and so we tentatively ascribe the $T_2$ to the line that dominates the $X^{-}$ emission in our experiments. Assuming that what limits the $X^-$ valley lifetime is the exciton scattering between valleys as shown in WSe$_2$\cite{Singh:2016a}, we now have the following simple picture that explains the unusually high circular polarization of the $X^-$ in MoS$_2$. For the $T_2$ in $K^+$, a change of the exciton valley would correspond to the formation of the complex shown in Fig.\ref{fig:fig4}(e), which is forbidden due to the Pauli principle. For the $T_1$ in $K^+$, a change in the exciton valley would lead to the formation of a triplet trion shown in Fig.\ref{fig:fig4}(f), which is likely to be unbound \cite{Roch:2019a} and, in any case, would recombine at a different energy than that of $T_1$ and $T_2$. To obtain emission coming from $T_1$ and $T_2$ in the $K^-$ valley  after $\sigma^+$ excitation would then require, in addition to the exciton intervalley transfer, a simultaneous spin (and momentum) flip of the extra electron. This simultaneous transfer of three carriers can explain the robustness of the valley degree of freedom of the negatively-charged trion in MoS$_2$; making it a very promising quantum state to store and manipulate information. More detailed experiments with time-resolution and tunable excitation energy could give in the future more insights into the role of excess energy and whether any significant valley relaxation occurs during the trion lifetime. \\

\begin{figure*}[htbp]
\includegraphics[width=1.05\textwidth]{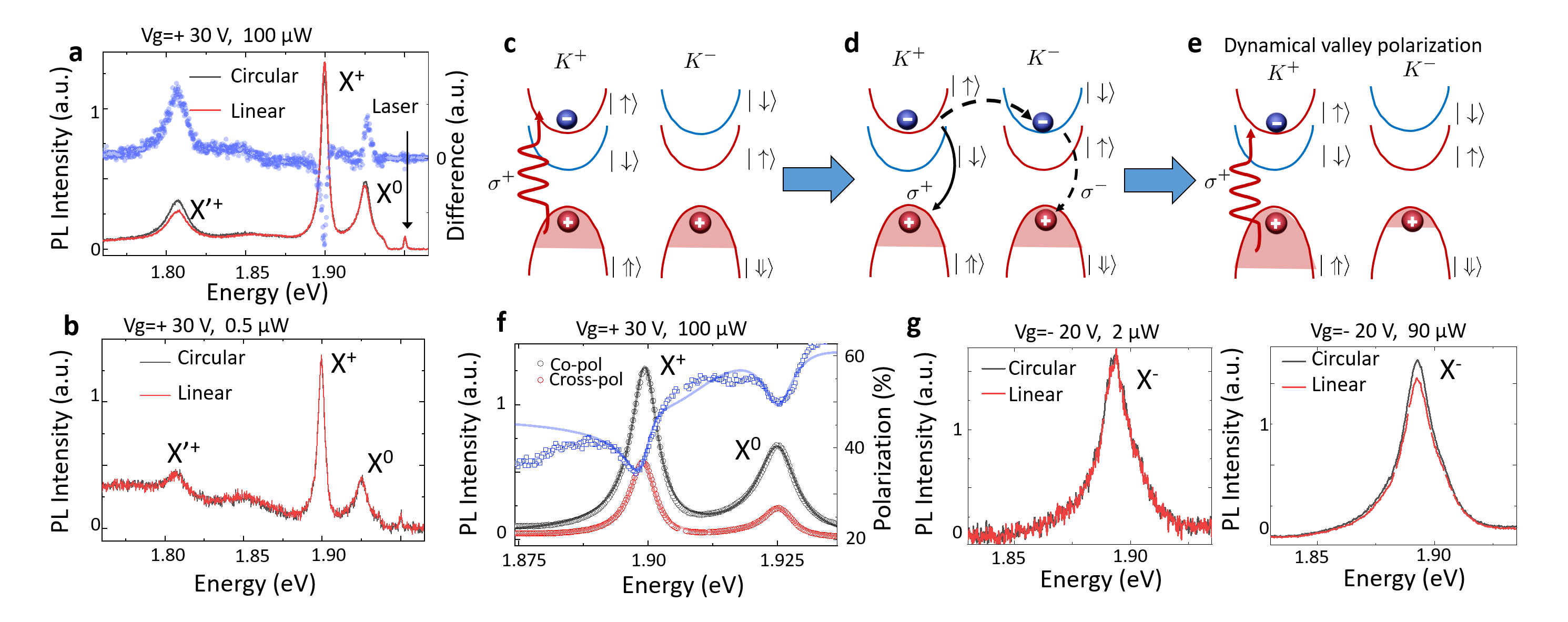}
\caption{\label{fig:fig5} (a) Total PL intensity under circular (black) and linear (red) excitation in the p-type regime at an excitation power of 100 $\mu$W. Also shown in blue dots the difference between these two curves. (b) Same as in (a) at a small excitation power of $2\;\mu$W. (c) Formation of a positively-charged trion in the $K^+$ valley following right-handed circular polarization. (d) A trion in the $K^+$ can either recombine radiatively or scatter to the opposite valley before recombination. (e) In steady state, the process (c) and (d) will lead to a spin and valley-polarization of the valence band. (f) Polarized PL emission under circular excitation and degree of circular polarization (blue dots). (g) Total PL intensity under circular (black) and linear (red) excitation in the n-type regime for an excitation power of $2\;\mu$W (left) and $90\;\mu$W (right).
}
\end{figure*}

We finally discuss the possibility of optically detecting a resident carrier polarization in monolayer MoS$_2$. As shown recently \cite{Robert:2021a}, efficient  optical valley pumping of resident carriers was demonstrated in n-doped WSe$_2$ and WS$_2$ monolayers with a continuous wave, circularly-polarized excitation. This manifests itself by a laser polarization-dependent PL intensity. Fig.\ref{fig:fig5}(a) shows the total PL intensity under circular (black) and linear (red) laser excitation in the p-type regime at $10\;\mu$W. Also shown is the difference between the two spectra (blue dots). We detect a small but measurable difference, with an excess of intensity for $X^0$ and $X^{'+}$ under circular excitation and a decrease in intensity for the $X^+$. Note that no difference is detected on the reflected laser intensity, excluding polarization-dependent absorption effects or laser power fluctuations. We have also checked that at low excitation power, no difference is observed in the PL emission between circular and linear excitation, as shown in Fig.\ref{fig:fig5}(b). We attribute this behaviour to the manifestation of a dynamical polarization of resident holes induced by the circularly-polarized cw excitation. Without light excitation and external magnetic fields, both $K^+$ and $K^-$ valleys are equally populated by holes. Upon $\sigma^+$ photoexcitation, excitons are photogenerated in the $K^+$ valley, where the electron in the exciton occupies the top conduction band in $K^+$. This electron can scatter to the $K^-$ valley, eventually recombining with a hole in $K^-$, resulting in a dynamical buildup of (hole)valley polarization. Another possible mechanism that can contribute to this build-up of valley polarization is illustrated in Fig.\ref{fig:fig5}(c)-(e). Here, an exciton in the $K^+$ valley binds to a hole in $K^-$ to form the positively-charged trion $X^+$ (Fig.\ref{fig:fig5}(c)). This trion can either  recombine radiatively in the $K^+$ valley, in which case no imbalance is created between the valleys, or it can scatter to the opposite valley before recombining. This is represented by the dashed arrows in Fig.\ref{fig:fig5}(d). In the latter case, photoexcitation has added a hole to the $K^+$ valley while substracting a hole in $K^-$. If the rate of photogeneration is large enough, and if the previously mentioned intervalley scattering mechanisms happens faster than the hole valley relaxation (which can be a very low process due to the significant spin-orbit splitting of the valence bands), this will lead to a larger hole population in $K^+$ valley, as shown in Fig.\ref{fig:fig5}(e). This valley-polarization of the valence bands can qualitatively explain the observations of Fig.\ref{fig:fig5}(a): if the $X^+$ formation is a bimolecular process (exciton binding followed by the capture of an additional hole of the opposite valley), then it is clear that upon circular excitation the $X^+$ formation is suppressed with respect to the case of linear excitation, resulting in a reduction in the $X^+$ PL. In contrast, an excess of the $X^0$ emission is expected since the trion formation rate, a process responsible for exciton dissapearance, will be smaller. The $X^{'+}$ intensity being larger under circular excitation is consistent with its assignation to a trion in which the additional hole is trapped by an in-gap state. The formation of this trion is in competition with the formation of the $X^+$, and so the reduction in $X^+$ formation rate can favour the formation of $X^{'+}$. \\

\noindent
In order to estimate the resident hole valley polarization, we use the fact that the $X^+$ emission energy depends on the position of the quasi Fermi-level of the opposite valley in the degenerate regime. This should produce a splitting of the $X^{+}$ emission in $\sigma^+$ and $\sigma^-$ polarization, and therefore an oscillation of the circular polarization inside the $X^+$ line. This is indeed observed in Fig.\ref{fig:fig5}(f), where the polarized PL emission under a circularly-polarized, $100\;\mu$W excitation is shown, together with the degree of circular polarization at each photon energy. The solid lines correspond to a fit with a double, asymmetrical peak. To fit the $\sigma^+$ and $\sigma^-$ components of the PL, we have kept all the parameters fixed except from the amplitudes and energy positions of each peak. We extract a splitting  of the $X^+$ of $\Delta E =\Delta E_{\sigma^+} -\Delta E_{\sigma^-} = 350 \pm 50\;\mu$eV. The positive sign is consistent with our picture: the emission energy of the $\sigma^+$ component should be blueshifted since the quasi-fermi level penetrate less into the $K^-$ valence band. We can roughly estimate the degree of valley polarization by using the previously estimated resident hole density of $p \sim 3 \times 10^{11}\;\mbox{cm}^{-2}$, which gives, for a temperature $T=20$ K and a hole effective mass of $m^*\approx 0.5 m_0$, a Fermi level of $\sim 490\;\mu$eV inside the valence band in the absence of light excitation. Since in each valence band the hole density is given by 

$$  p^{\pm} = \frac{k_B T m^*}{2\pi \hbar^2} \ln \left( 1+e^{(E_v-E_F^{\pm})/k_B T} \right) $$

\noindent
with $E_F^{\pm}= 490 \pm 350 \;\mu$eV, we estimate a hole valley polarization of $\sim 15 \;\%$. This is significantly smaller than the resident carrier valley polarization of $80\;\%$ achieved in n-doped WSe$_2$ monolayers, and consistent with the much smaller variations of PL intensity under circular and linear excitation observed here in monolayer MoS$_2$. The smaller spin-orbit splitting of the valence band in Mo-based TMDs may be at the origin of this smaller steady-state hole valley polarization. \\

\noindent
We note that also a small splitting of the $X^0$ is responsible for a rapid oscillation of the circular polarization inside the exciton line, which is absent in the neutral regime (Fig.\ref{fig:fig2}(a)). Indeed the valley polarization of the valence band should give rise to a valley-dependent $X^0$ binding energy. We have found that the asymmetry in the high-energy side of the $X^0$ line is different for the $\sigma^+$ and $\sigma^-$ components, making a reliable determination of the splitting difficult.  Finally, we show in Fig.\ref{fig:fig5}(g) that an optical signature of the valley polarization of resident electrons is also observed in the n-type regime for a sufficiently large excitation power. Here, an easy estimation of the valley polarization is not possible due to the broad emission of the negatively-charged trions. Importantly, these results demonstrate that the dynamical valley polarization and spin-dependent PL emission is not restricted to tungsten-based TMD monolayers.\\

In summary, this work brings new elements for the understanding on the different mechanics that may influence the robustness of the valley polarization of quasiparticles in TMD monolayers and the dynamical valley polarization of resident carriers. We have demonstrated p and n-type electrostatic doping in MoS$_2$ monolayers and study the valley polarization of several excitonic complexes. Several of these complexes have never been reported before. We find large steady-state circular polarization upon circular excitation for all these excitonic complexes.  The inverted band ordering  between the single particle and the exciton picture is a particular feature of MoS$_2$ and makes negatively-charged trions in this material very promising candidates for future valleytronic applications, exhibiting remarkably high circular polarization in PL emission. We have also demonstrated the signatures of an optically-induced dynamical valley polarization of resident electrons and holes,which can be achieved with continous wave circularly-polarized excitation.This work is thus an important step towards the development of valleytronic devices based on TMD MLs. Indeed, both highly-polarized trions and valley polarized resident carriers influence the transport and the optical properties of TMDs and may be detected and manipulated in experiments such as the valley Hall effect \cite{Mak:2014a}.\\

\indent \textit{Methods.---} 
Silicon dioxide layers were thermally grown on a 3-inch diameter silicon wafer (100, p-type 5-10 ohm cm) in a tube furnace (dry oxidation conditions). The thickness of the oxide was measured to be 90.1$\pm$1.2 nm using ellipsometry  ( AutoSE spectroscopic ellipsometry, Horiba). Next an ebeam-based lithographic/lift-off  microfabrication process was performed involving two thermal evaporation metallization steps to obtain large (500 µm × 500 µm) contact/bonding pads having a Cr/Au thickness of 5/300 nm and long, thin (1 µm or 3 µm) conducting lines having a thickness of 5/50 nm which lead up to the target area for the ML deposition. The thickness of the metallizations was verified using surface profiling ( DektakXT - Bruker, USA). The wafers were diced to obtain individual chips measuring 8 mm by 8 mm. For the gate-dependent PL measurements, a Keithley 2400 sourcemeter was used with the ground connected to the bottom graphite flake. \\

\indent \textit{Acknowledgements.---} 
F.C, S. P, and F. S ackowledge the Grant "SpinCAT" No. ANR-18-CE24-0011-01. F.C. would like to thank Romain Grasset for help with the electrical setup.
.  K.W. and T.T. acknowledge support from the Elemental Strategy Initiative
conducted by the MEXT, Japan (Grant Number JPMXP0112101001) and  JSPS
KAKENHI (Grant Numbers 19H05790, 20H00354 and 21H05233). S.A acknowledges that part of the work was funded by the French RENATECH network. S.A would like to thank Guillaume Cochez, François Vaurette, and Annie Fattorini for help with the thermal oxidation, the ebeam lithography, and the metallization.

%

\end{document}